\documentclass[nofootinbib,notitlepage,superscriptaddress,10pt,aps,pra,twocolumn]{revtex4-1}

\usepackage[utf8x]{inputenc}
\usepackage{amsmath}
\usepackage{amssymb}
\usepackage{bbm}
\usepackage{graphicx}
\usepackage{mathptm}
\usepackage{color}

\begin{document}

\title{Quantum-spacetime scenarios and soft spectral lags of the remarkable GRB130427A}

\author{Giovanni Amelino-Camelia$^a$, Fabrizio Fiore$^{b}$, Dafne Guetta$^{b,c}$ and Simonetta Puccetti$^{b,d}$\\
{\small{$^a$\it{Dipartimento di Fisica, Sapienza Universit\`a di Roma and INFN, Sez.~Roma1, P.le A. Moro 2, 00185 Roma, EU}}}\\
{\small{$^b$\it{INAF-Osservatorio Astronomico di Roma, via Frascati 33, I00040 Monteporzio  Catone, Italy}}}\\
{\small{$^c$\it{Department of Physics, ORT Braude, Snunit 51 St. P.O.Box 78, Karmiel  21982, Israel}}}\\
{\small{$^d$\it{ASI Science Data Center, via Galileo Galilei,00044 Frascati Italy}}}}

\begin{abstract}
\noindent
We process
the Fermi LAT data on GRB130427A using the Fermi Science Tools,
and we summarize some of the key facts that render this observation truly remarkable, especially
concerning the quality of information on high-energy emission by GRBs.
We then exploit this richness for a search of spectral lags, of the type that has been recently
of interest for its relevance in quantum-spacetime research.
We do find some evidence of systematic soft spectral lags: when confining the analysis
to photons of energies greater than 5 GeV there is an early hard development
of minibursts within this long burst. The effect turns out to be well characterized quantitatively by a linear dependence,
within such a miniburst, of the detection time on energy. With the guidance of our findings for GRB130427A
we can then recognize that some support for  these features is noticeable also in earlier
Fermi-LAT GRBs, particularly for the presence of hard minibursts  whose onset is
 marked by the highest-energy photon observed for the GRB.
 A comparison of these features for GRBs at different redshifts
 provides some encouragement for a redshift
dependence of the effects of the type expected for a quantum-spacetime interpretation, but other aspects
of the analysis appear to invite the interpretation
as  intrinsic properties of GRBs.
\end{abstract}

\maketitle

\baselineskip11pt plus .5pt minus .5pt

\section{Introduction and motivation}
The study of Gamma-Ray Bursts (GRBs) has been for some time one of the main themes
of astrophysics research, and over the last 15 years it became also of interest
for research on quantum gravity. These more recent developments look at GRBs as signals,
and hope to decode properties of quantum
spacetime from the implications of spacetime quantization for how such signals propagate
from the distant source to our telescope. Even tiny quantum-spacetime effects modifying
the structure of the signal could cumulate along the way, as the signal travels over
cosmological distances.

As we here contribute to assess, the remarkable very recent observation
of GRB130427A is bound to teach us a lot about the astrophysics of GRBs and has
the potential to also empower some of the quantum-spacetime studies of GRBs.

This GRB130427A was extremely powerful, also thanks to the fact that it was
among the nearest long GRBs observed. We here take the perspective
that GRB130427A is an opportunity to look at a long GRB in ``high resolution", allowing
us to notice features which could not be noticed in previous GRBs.

In this study we test the potentialities of using GRB130427A in this way by focusing
on a much studied class of effects, motivated by quantum-spacetime research.
These are effects such that a group of photons ideally emitted all in exact simultaneity
at some distant source should reach our telescope with systematic spectral lags.

In the next section, besides discussing some aspects of the exceptionality of GRB130427A and
the procedures we followed for retrieving Fermi-LAT data on this GRB,
we briefly describe the quantum spacetime picture that inspired this GRB phenomenology.
We focus in particular on the most studied model of this class of quantum-spacetime effects,
which predicts spectral lags whose magnitude grows linearly with energy and depends on
redshift in a characteristic way.

Then in Section III we look at GRB130427A Fermi-LAT data from the perspective of these linear spectral lags.
We do not find any evidence of such spectral lags on the low side of the LAT energy range,
but we notice that if one restricts attention to GRB130427A photons with energy in excess of 5 GeV
there is some rather robust evidence of spectral lags, specifically linear soft spectral lags.
This feature is even more pronounced if one restricts attention to GRB130427A photons with energy
in excess of 15 GeV.
And in the course of characterizing these spectral lags we also stumble upon a feature
which might be labeled ``hard minibursts":  if one restricts attention to GRB130427A photons with energy in excess of 5 GeV
a significant fraction of them is found to be part of bursts that last much shorter than the time scale set
by the  overall duration of the GRB130427A  signal in the LAT. And our analysis finds
that the first photon in such hard minibursts always is the highest-energy photon in the miniburst.

Section IV is where we try to use our findings for GRB130427A as guidance for noticing
features in data available on previous GRBs.

In the closing Section V we offer a perspective on our findings and comment on various scenarios
for their interpretation.

\section{Preliminaries on GRB130427A and quantum spacetime}
\subsection{The remarkable GRB130427A and Fermi-LAT data}

At 07:47:06 UT on 27 April 2013, Fermi-LAT detected\cite{zhu} high energy
emission from GRB 130427A, which was also
detected\cite{kienli} at lower energies by Fermi-GBM
and by several
other telescopes including Swift\cite{maselli}, Konus-Wind\cite{golene},
SPI-ACS/INTEGRAL\cite{pozanenko}
and AGILE\cite{verrecchia}.

\begin{figure*}[htbp!]
\includegraphics[width=1.45 \columnwidth]{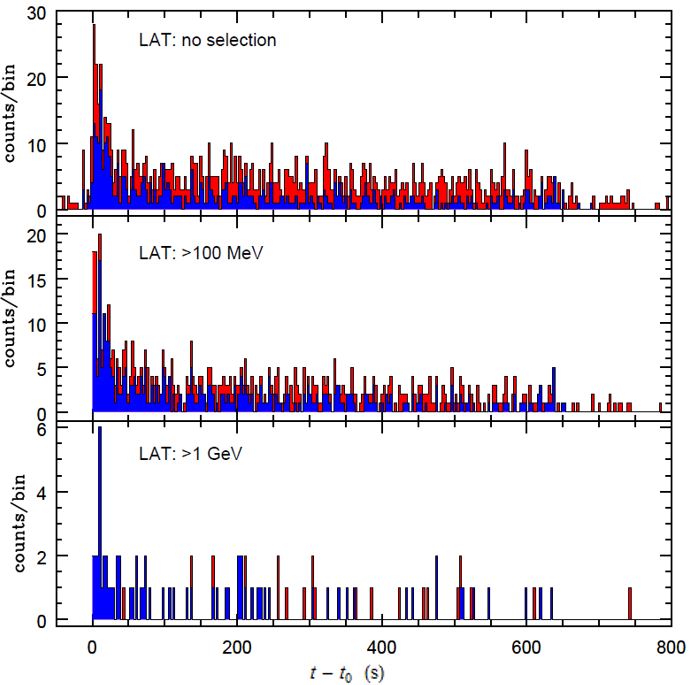}
\caption{Fermi-LAT light curves for GRB130427A (bin size of 3 seconds).
Photons selected within 10 degrees
of the GRB130427A optical position. The contribution from the subset of photons
within 3 degrees  of the GRB130427A optical position is shown in blue.}
\end{figure*}

From these early (and however sketchy) reports it emerges that
GRB130427A is a record setter in many ways. From the point of view of
high-energy astrophysics and fundamental physics, relevant for the
study we are here reporting, particularly significant is
the fact that GRB130427A is the new record holder as the
highest-fluence LAT-detected GRB.  And the GRB130427A signal contains
a $\simeq 95 GeV$ photon, which establishes the new record for a GRB,
by a very wide margin (all previous Fermi-LAT GRB-photons had energy
lower than 35 GeV).

Indications of an extreme fluence for this GRB is also confirmed at
lower energies by the SPI-ACS detector onboard INTEGRAL and by Swift.
And Swift reports an extremely luminous X-ray afterglow.

Looking back at GRB catalogues one can tentatively estimate that such an
exceptional GRB is only observed not more frequently than
once in a quarter century (a good benchmark\cite{grb881024}, on the basis of fluence,
might be GRB881024), and the opportunity is exciting since the quality
of our telescopes made so much progress over this last quarter of a
century.

The long list of records is surely in part due to the fact that
GRB130427A was a rather near long GRB.  The redshift was determined\cite{flores,levan,xu} to
be $0.3399 \pm 0.0002$, so we probably got to see from up close (in ``high
definition") a GRB which already intrinsically was rather powerful.

For the purposes of the analysis we are here reporting, we retrieved
LAT Extended Data, in a circular region centered on the optical
position\cite{elenin} R.A.$=$11:32:32.84,Dec.$=+$27:41:56  from
the Fermi LAT data server.

We prepared the data files for the analysis using  the LAT
ScienceTools-v9r31p1 package, which is available from the {\it Fermi}
Science Support Center.
An overall characterization of
the data we retrieved is offered in Fig.1, which shows the Fermi-LAT light curve of GRB130427A.
Following
the photon selection suggested by the Fermi team for GRB analysis, we selected all the events of ``P7TRANSIENT'' class or
better, in the energy range 100 MeV-300 GeV and with the zenith angle
$\le$ 100 deg.  ``P7TRANSIENT'' class contains lower quality photons,
the "transient" class events. These cuts are optimized~\cite{acke} for transient
sources for which the relevant timescales are sufficiently short that
the additional residual charged-particle backgrounds are much less
significant. We selected photons in a circular region
centered on the GRB130427A optical position and with a maximum radius
of 10 degrees which ensures that all GRB photons ('FRONT' and 'BACK',
as described in Ref.\cite{atwood2009}) are considered.
Nevertheless, since our study is mainly focused on photons with
energy greater than 5 GeV, and because the LAT Point Spread Function
is energy dependent, we have decided to use also a more conservative
source extraction region with a maximum radius of 3 degrees, which
corresponds to about $60\%$  and about $90\%$ of the LAT Point Spread Function at
energy greater than 5 GeV for the P7TRANSIENT and P7SOURCE class~\cite{acke2012},
respectively. More on this aspect in Subsection III.C.

The times and energies of detection of
 GRB130427A photons with energy in excess of 5 GeV are shown
in Fig.2.

\begin{figure}[htbp!]
\includegraphics[width=0.89 \columnwidth]{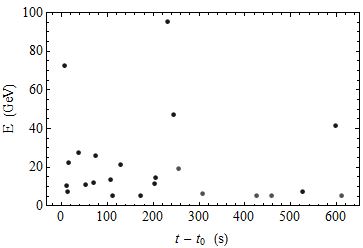}
\caption{Times and energies of detection of GRB130427A photons with energy in excess of 5 GeV,
selected within 10 degrees
of the GRB130427A optical position. Darker points are for photons observed from within
3 degrees  of the GRB130427A optical position.}
\end{figure}

\subsection{A scenario for quantum spacetime}
A rich phenomenological program was developed over the last decade on the basis
of results
for models of spacetime quantization suggesting
that (see, {\it e.g.}, \cite{grbgac,gampul,urrutia,gacmaj,myePRL})  it is possible for the quantum properties of spacetime to introduce small violations of the special relativistic properties of classical spacetime.
A particularly valuable opportunity
for phenomenology is provided by pictures of quantum spacetime in which
 the time needed for a ultrarelativistic particle\footnote{Of course the only regime of particle propagation
 that is relevant for this manuscript is the ultrarelativistic regime, since photons have no mass and
 the neutrinos we shall consider toward the end of the manuscript
 have energies such that their mass is completely negligible.}
to travel from a given source to a given
detector is  $t=t_0+t_{QG}$.  Here $t_0$ is the time that would be predicted
in classical spacetime,
while $t_{QG}$ is the contribution to the travel time due to quantum properties of spacetime.
 For energies much smaller than $M_{QG}$,  the characteristic scale  of  these quantum-spacetime effects,
 one expects that at lowest order $t_{QG}$ is given by  \cite{jacobpiranDELAY}:
\begin{equation}
t_{QG} = - s_\pm \, \frac{E}{M_{QG}} \frac{D(z)}{c}  ,
\label{main}
\end{equation}
where
\begin{equation}
D(z) = \frac{c}{H_0} \int_0^z d\zeta \frac{(1+\zeta)}{\sqrt{\Omega_\Lambda + (1+\zeta)^3 \Omega_m}}  .
\nonumber
\end{equation}
Here the information cosmology gives us on spacetime curvature is coded
in the denominator for the integrand
in $D(z)$, with $z$ being the redshift and $\Omega_\Lambda$, $H_0$ and $\Omega_0$ denoting, as usual,
respectively the cosmological constant, the Hubble parameter and the matter fraction.
The ``sign parameter" $s_\pm$, with allowed values of $1$ or $-1$, as well as the scale $M_{QG}$  would have to be determined
experimentally. We must stress however that most theorists
favor naturalness arguments suggesting that $M_{QG}$ should take a value
that is
rather close to
the ``Planck scale"
$M_{planck} =\sqrt{{\hbar c^5}/{G_N}} \simeq 1.22 \cdot 10^{19}GeV~.$

 The picture of quantum-spacetime effects
summarized in (\ref{main}) does not apply to all quantum-spacetime models. One can evisage quantum-spacetime pictures
that do not violate Lorentz symmetry at all, and even among the most studied quantum-spacetime pictures
that do violate Lorentz symmetry one also finds
variants producing (see, {\it e.g.}, \cite{grbgac,mattiLRR,gacsmolin})  features analogous to (\ref{main}) but with
the ratio $E/M_{QG}$ replaced by its square, $(E/M_{QG})^2$, in which case the effects
would be much weaker and practically undetectable at present.

It is important that  some quantum-spacetime models allow for laws roughly of the type (\ref{main})
to apply differently to different types of particles.
For example, in the so-called ``Moyal non-commutative spacetime", which is one of the most studied  quantum spacetimes, it is remarkably found \cite{szabo} that the implications of spacetime quantization for particle propagation end up depending on the standard-model charges carried by the particle and its associated coupling to other particles.

But let us stress, since it will be relevant later on in our discussion, that, while particle-dependent effects would indeed
not be surprising, we know of no model that predicts an abrupt onset of such effects: the quantum-spacetime model building so far available offers candidates for  particle-dependent effects, but in none of the known models the effects are exactly absent at low energy and then switch on to full strength above some threshold value of energy.
Formula  (\ref{main}) illustrates the type of mechanisms that are known in the quantum-spacetime literature, with the effects confined to the ultraviolet by the behaviour of analytic functions introducing energy dependence such that the effects are small at low energies and become gradually stronger at higher energies.

This point is important since there is an apparently robust limit at $M_{QG} > 1.2 M_{Planck}$
derived in Ref.\cite{fermiNATURE} (also see Refs.\cite{gacsmolin,unoSCIENCE,ellisPLB2009,hessREVIEW,nemiro})
applying two different techniques to GRB090510 data: for one technique photons with energy below 5 GeV play the primarily role
 and is based primarily on statistical properties of the arrival-time vs energy GRB090510 data on such photons, whereas  the
 other technique exploits mainly the good coincidence of detection times between a single 31 GeV photon and a very short
 duration burst of GRB090510 lower-energy photons.
It should also be noticed that at energies higher than the range of energies we shall consider for GRB130427A the
implications of Eq.(1) have been studied through observations of Blazars by ground telescopes such as MAGIC and HESS,
and this also produces \cite{hessREVIEW,magic,hessPRL2008} significant bounds on $M_{QG}$ at a level just below
the Planck scale. In light of this the initial objective of  our study was to establish whether the
unusual quality of GRB130427A data could be used to study the effects of Eq.(1) with sensitivity even better than
the Planck-scale level. Instead we found, as shown already in the next section,
evidence of features stronger than suggested by Eq.(1) with $M_{QG} \simeq M_{Planck}$
in the structure of arrival times versus energy for GRB130427A
photons, which in turn (also in light of the points summarized in this subsection)
 raises several challenges of interpretation.

\section{Soft spectral lags for minibursts within GRB130427A}

\subsection{Spectral lags for minibursts within GRB130427A}
As evident from what we already summarized in Subsection II.B, part of our interest in GRB130427A
originates from it being an opportunity for looking, at higher energies than in previous GRBs,
for spectral lags of the type predicted
for GRBs on the basis of (\ref{main}).

Looking from this perspective at the GRB130427A  data summarized in Fig.1, including only photons with energy
greater than 5 GeV, one cannot fail to notice a very clear hint of systematic spectral lags.
This is highlighted in Fig.3: if we assume the validity of (\ref{main}), with  $M_{QG} \simeq  M_{planck}/25$
and $s_\pm =1$, then for each of the four highest-energy photons in GRB130427A we  find
at least one more photon
(in one case two other photons) which could have been emitted simultaneously with it.
And we here actually mean simultaneously, not near-simultaneously: for example for the
impressive three high-energy photons on the central line of Fig.3 the analysis shows that,
within experimental uncertainties, the times of arrival are consistent with exact simultaneity of emission
if indeed one assumes (\ref{main}), {\it i.e.} the differences in times of arrival
between photons on such lines could be fully accounted for by (\ref{main}) starting from simultaneous emission.

\begin{figure}[htbp!]
\includegraphics[width=0.89 \columnwidth]{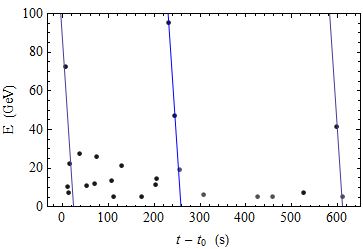}
\caption{The three parallel straight lines here shown identify cases where photons on one such worldline could
 have been even emitted simultaneously (up to experimental errors) if Eq.(1) holds at the relevant energies.}
\end{figure}

The central line  of Fig.3 is impressive because it suggests that 3 of the top-9 most energetic photons
in GRB130427A might have been emitted in exact simultaneity at the source, if (\ref{main}) is correct at those energies.
And the other two lines, exactly parallel to the central one, suggest that this might not have been accidental:
for all the top-4 most energetic
photons in GRB130427A (the 4 photons with energy greater than 40 GeV) one finds at least a photon candidate for simultaneous
emission according to (\ref{main}) within our rather limited
sample of the 23 GRB130427A-photons with energy greater than 5 GeV.

Another way to visualize the content of  Fig.3 is to
 consider the 7 points on those 3 lines and fit the values of  the time translations needed to superpose the 3 lines.
The result of this fit is shown in Fig.4, where we also show the single line which
then best fits the 7 points.
For the slope one finds $(-2.81 \pm 0.27) GeV/s$ which in the spirit of Eq.(1) could be described
as the case with $s_\pm =1$ and $M_{QG} =(3.8 \pm 0.4)10^{-2}M_{planck}$.

The chi squared of the fit is of 5.8 which might not look too impressive
since our 7 data points were also used to determine the slope, besides determining the translations needed to superpose
the lines. But actually the quality of this fit is surprisingly good if one takes into account that
the implicit assumption behind this fit is that  photons sharing a line in Fig.3
should have been emitted in exact simultaneity at the source. We are contemplating such an unrealistically
 rigid constraint just so that the strength of our case can be easily perceived.

\begin{figure}[htbp!]
\includegraphics[width=0.89 \columnwidth]{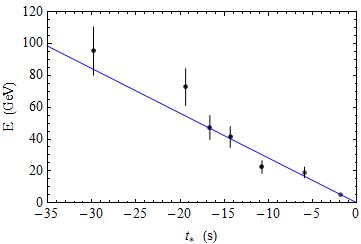}
\caption{Here we show the results of a fit that superposes the three parallel lines in Fig.3. The seven points
here shown are the ones on those parallel lines of Fig.3.}
\end{figure}

\vskip -3cm

\subsection{Supporting evidence from a correlation study}

The observation visualized in Figs.3 and 4 is evidently noteworthy but it
is not easy to quantify from it a statistical significance of the feature we are contemplating.
As a way to compensate for that we have performed a correlation study.
 We essentially studied the frequency of occurrence
of spectral lags in our sample as a function of a minimum photon energy and without any prior assumption on high energy photons.
This ultimately gives us an alternative way for determining the slope of lines
such as those in Fig.3 that fit well the spectral lags present in the data.
And this comes with an easy estimate of statistical significance.

\begin{figure}[htbp!]
\includegraphics[width=0.71 \columnwidth]{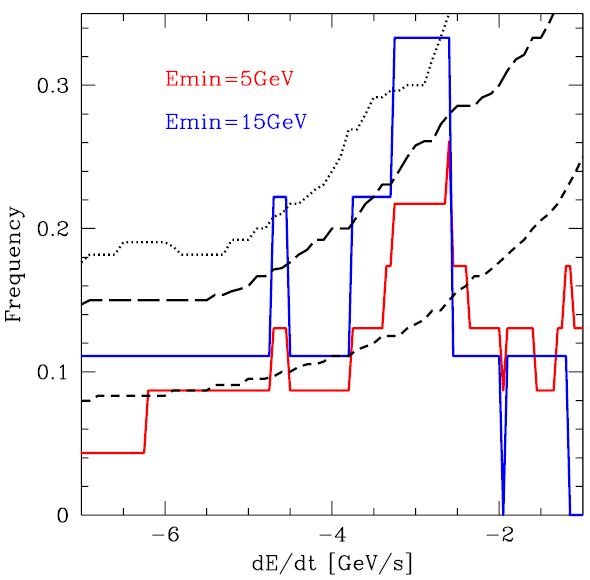}
\caption{Here shown is a study
of the frequency of occurrence of spectral lags in our sample
as a function of a minimum photon energy and without any prior assumption on high energy photons.
The red (respectively blue) solid line describes the frequency of occurrence of a certain
spectral-lag slope within the sample of GRB130427A events with energy grater than 5 GeV
(respectively 15 GeV).
Dotted/dashed lines  represent  the $10\%$, $1\%$ and $0.23\%$ probability that the given
frequency of occurrence of spectral lags is a chance occurrence.}
\end{figure}

Our result for this correlation study is summarized in Fig.5, which gives a result
in perfect agreement with the message conveyed in the Figs.3 and 4 of the previous subsection,
including the determination of the slope $dE/t$ of about $-3$ in GeV per second.
Dotted/dashed lines in Fig.5 represent  the  $10\%$, $1\%$ and $0.23\%$  probability that the given
frequency of occurrence of spectral lags is a chance occurrence.
The lines have been calculated using 10000 Monte Carlo simulations of photon energy versus time
realizations assuming the same photon energy distribution as the data.
Fig.5 shows
that the probability of chance occurrence of spectral lags of about 3 GeV/second is $\lesssim 0.23\%$
for an energy threshold of $15 GeV$ and just below the  $1\%$ level for an energy threshold of 5 GeV.

This comparison between our correlation study with a 5GeV lower cutoff and our correlation study with a 15GeV
lower cutoff also suggests that the feature we are exposing emerges at high energies. This after all is also
the message of Fig.3 where one immediately sees a pattern of spectral-lag correlations among the highest-energy
photons and between the highest energy photons and some lower-energy photons, whereas such correlations
among pairs of photons both with energy smaller than 15GeV are not clearly visible (if at all present).
We confirm this message with the study visualized in Fig.6 which is of the same type reported in Fig.5 but focuses
on GRB130427A photons with energy between 1 and 5 GeV: contrary to what shown in Fig.5 we find no significant
signal in the study shown in Fig.6.

\begin{figure}[htbp!]
\includegraphics[width=0.71 \columnwidth]{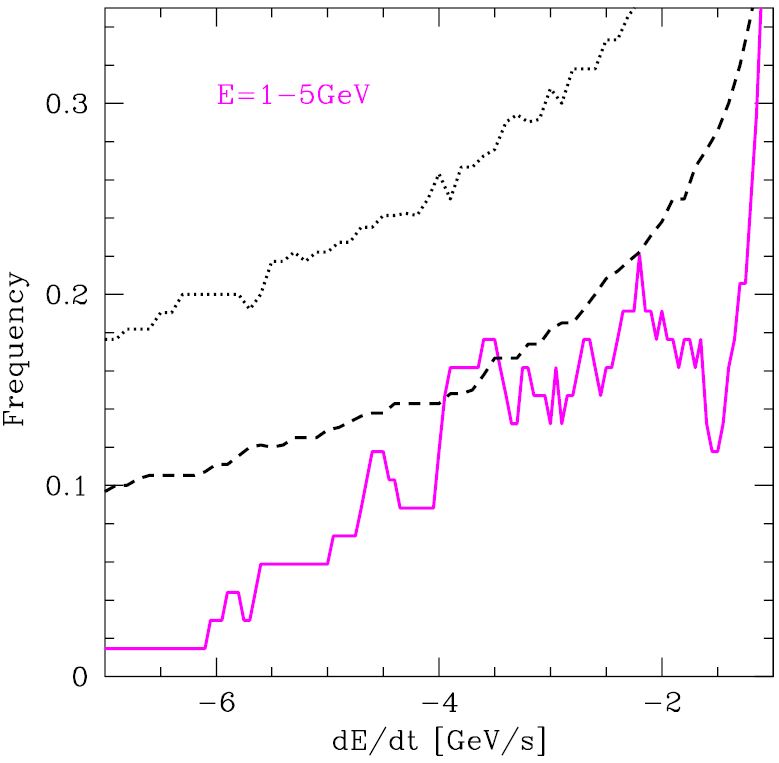}
\caption{Here shown are the results of a study of the type already described in the previous Fig.5,
but now restricting the analysis to photons with energy between 1GeV and 5 GeV. Again
the dotted line and the dashed line estimate respectively the $1\%$ and $10\%$ probability that the given
frequency of occurrence of spectral lags is a chance occurrence.}
\end{figure}

\newpage

\subsection{Two mechanisms and background issue}

The fact that evidence of systematic spectral lags appears to be noticeable
at particularly high energies
may suggest a threshold behaviour for the emergence of new properties, such as in the case
of laws of propagation in quantum spacetime governed by (\ref{main}) but only above a certain
threshold value of the energy of the particles.
And/or it may
 suggest that in some time intervals during the burst two rather different mechanisms coexist, one
affected strongly by the spectral lags and the other one unaffected by them (or affected in much weaker
way).
Concerning this latest possibility by looking at the data points
in Fig.3 one finds encouragement for contemplating the possibility that
during the first 190 seconds after the trigger of GRB130427A a second mechanism dominates,
with the soft spectral lags which we are exposing being only a subdominant feature.
Fig.3 also suggests that after that 190-second window
within GRB130427A the roles might be reversed, with the soft spectral
lags being then a strongly characterizing feature.

We quantify this in Fig.7 by performing a spectral-lag correlation study completely
analogous to the one reported in Fig.5 but now focusing on a 450-second interval, from 190 seconds
to 640 seconds after the trigger. And what we find in Fig.7 appears to provide strong validation
to the hypothesis that at later stages in GRB130427A our soft-spectral-lag feature acquires
a more dominant role.
Comparing Fig.7 to Fig.5 one sees clearly that after the first 190 seconds
the evidence of spectral lags with a slope of about -3 is significantly stronger.
In particular, when focusing on times after the first 190 seconds also the study
of correlations taking into account all photons with energy greater than 5 GeV
produces an outcome whose probability of chance occurrence is significantly less than $0.23\%$.

\begin{figure}[htbp!]
\includegraphics[width=0.71 \columnwidth]{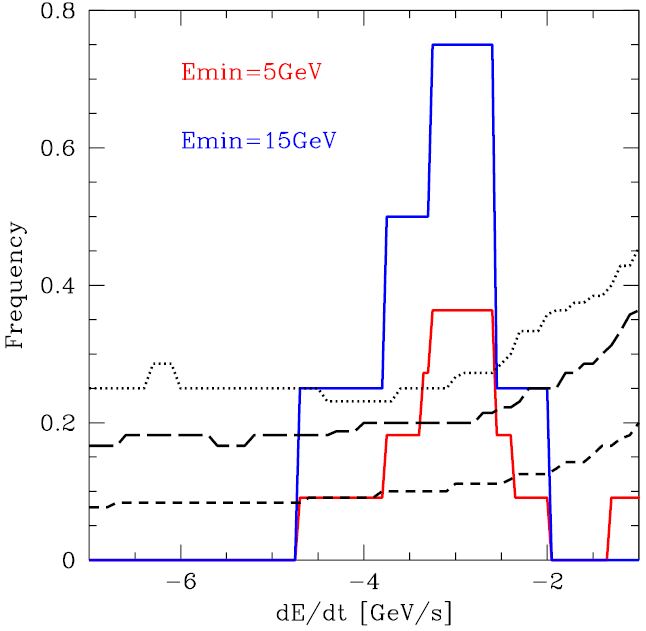}
\caption{Here shown are the results of a study of the type already described in the previous Fig.5,
but now restricting the analysis to the time interval from 190 seconds
to 640 seconds after the trigger.
Dotted/dashed lines  represent  again the $10\%$, $1\%$ and $0.23\%$ probability that the given
frequency of occurrence of spectral lags is a chance occurrence.}
\end{figure}

Considering the potential significance of the findings reported in Fig.7
it is appropriate here to pause for an assessment of the possibility that 
background events might be affecting
the rather striking evidence we are gathering.
In this respect it is noteworthy that two events among those
that contribute to the result of Fig.7 were observed at more than
3 degrees (but less than 10 degrees, see Section I) off the optical position
of GRB130427A. We must therefore attribute to them a somewhat higher risk
(higher than for those observed within 3 degrees of GRB130427A) of being background events.
These two events are the 19 GeV event visible in the bottom central part of Fig.3
at about 250 seconds after trigger and the 5.2 GeV event visible in the bottom right part of Fig.3
at about 610 seconds after trigger.

In Figs.8 and 9 we offer a quantification of how different the outlook of our analysis would
be when considering the possibility of excluding from it 
one of these two events. Comparing Fig.8 to  Fig.7 one sees
that renouncing to the 5.2 GeV event has tangible but not very significant implications
for our analysis. Comparing  Fig.9 to  Fig.7 one sees
that renouncing to the 19 GeV event does have rather significant implications
for our analysis, but our findings would be noteworthy even without
the 19 GeV event.

In light of this special role played by the 19 GeV event,
we have estimated  the probability of a background event of energy between 15 and 30 GeV
within the 450-second window considered for Fig.7. We find that this probability is of about $15\%$
 by both modeling the background and
using data available in a temporal and directional neighborhood of GRB130427A.
It appears legitimate to also notice that if such a 19 GeV event was truly background
and therefore unrelated to GRB130427A it could have come at any time within that 450-second
window, but it turned out to contribute so strongly to the significance of our analysis
only because it happened to be an excellent match for the central straight line in Fig.3.
In order for it to be such a strong contributor to the significance of our analysis it
had to be detected within a temporal window around that straight line of about 5 seconds,
so we are dealing with a $15\%$ probability overall but only a $\sim 0.17 \%$ probability
of a background event contributing that strongly to our case.

As stressed above, if we were to exclude the 19 GeV event we would still be left
with the noteworthy outcome shown in Fig.9. And evidently
the blue line in Fig.9 is independent
of the contribution by the 5.2 GeV event. We can nonetheless contemplate the possibility
that both the 19 GeV event and the 5.2 event might have been both background events which
 happened to be detected just at the right times to improve the outlook of our analysis.
 We estimate the ``conspiracy factor" of this hypothesis at no more than one chance in $10^4$.


\begin{figure}[htbp!]
\includegraphics[width=0.71 \columnwidth]{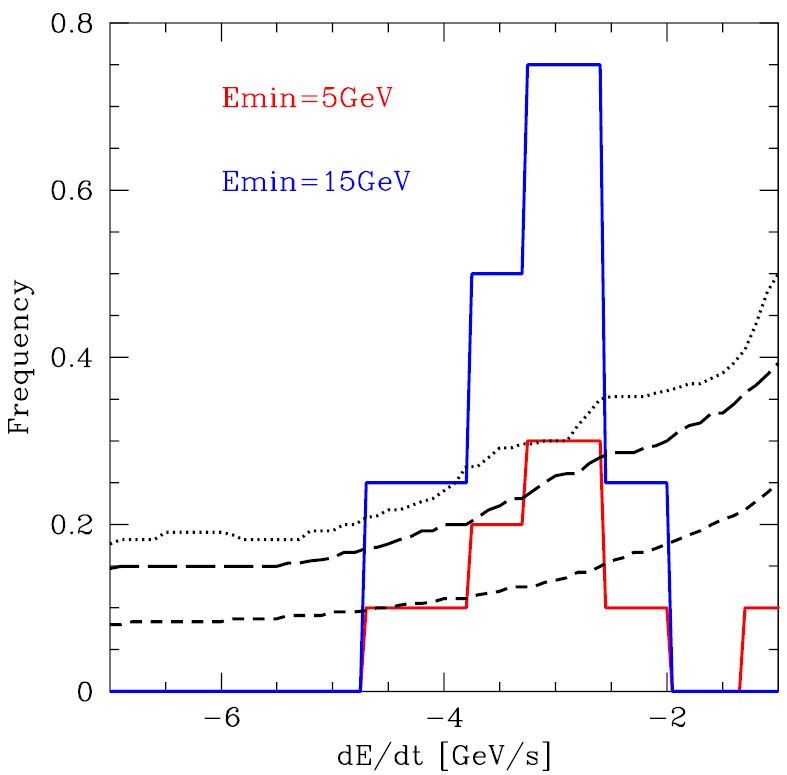}
\caption{Here shown are the results of a study exactly of the type already described in the previous Fig.7
(including the restriction to the time interval from 190 seconds
to 640 seconds after the trigger), but now excluding from the analysis
a 5.2 GeV event from about 610 seconds after trigger.
Dotted/dashed lines  represent  again the $10\%$, $1\%$ and $0.23\%$ probability that the given
frequency of occurrence of spectral lags is a chance occurrence.}
\end{figure}

\begin{figure}[htbp!]
\includegraphics[width=0.71 \columnwidth]{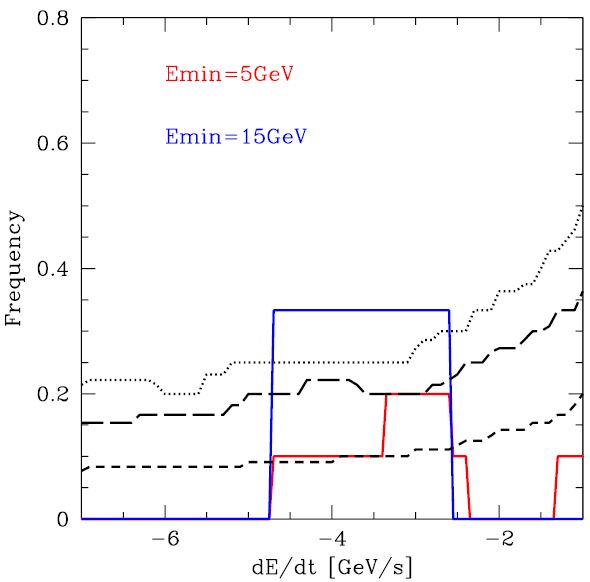}
\caption{Here shown are the results of a study exactly of the type already described in the previous Fig.7
(including the restriction to the time interval from 190 seconds
to 640 seconds after the trigger), but now excluding from the analysis
a 19 GeV event from about 250 seconds after trigger.
Dotted/dashed lines  represent  again the $10\%$, $1\%$ and $0.23\%$ probability that the given
frequency of occurrence of spectral lags is a chance occurrence.}
\end{figure}

\section{Further evidence, interpretation, and redshift dependence}
This study got started looking at the remarkable GRB130427A as an opportunity
to study spectral lags of the type expected from (\ref{main})
with sensitivity even better than
the Planck-scale level.
In the previous section we did stumble upon evidence, using indeed GRB130427A data,
that indeed some systematic spectral lags are found, but only when restricting the search
to high energies in an appropriate way, and only assuming a value of $M_{QG}$ significantly smaller
than $M_{Planck}$.
As stressed in Subsection II.B the need of restricting the analysis to energies higher
than some threshold value was not expected by the quantum-spacetime researchers
whose work motivates (\ref{main}).
Moreover, as also stressed in Subsection II.B, there were previous tests of Eq.(1) establishing that $M_{QG}$
should be higher
than $M_{Planck}$.
Plausible hypothesis appear to be then that either that Eq.(1) holds only in some energy range
(which however would be disfavoured by all theory works so far produced on this subject)
or the origin of the features we noticed is not connected to spacetime properties but rather reflects
properties of GRB sources.

We must also stress  that so far we have not probed much of the
structure of (\ref{main}): we did have encouraging results from applying the linear dependence
on energy of (\ref{main}), but of course having analyzed so far a single burst
we have in no way probed the redshift dependence coded in (\ref{main}).

Since we set off with the goal of testing (\ref{main}) from a quantum-spacetime perspective
but our findings led us to also
consider other possibilities, it is useful to start this section by reassessing the results
reported in the previous section in more model-independent manner.

What we did find so far is some evidence of systematic soft spectral lags, present when confining the analysis
to photons of energies greater than 5 GeV, and significantly more pronounced when the analysis is confined
to photons of energies greater than 15 GeV. These spectral lags
take the shape of an early hard development
of minibursts within a long burst. The quantitative aspects turn
out to be well characterized quantitatively by a linear dependence,
within such a miniburst, of the detection time on energy.
In particular, in each miniburst the first photon is the highest energy one.

And we strikingly found that each of the top 4 most energetic photons
of GRB130427A happened to be connected to at least another photon
with energy greater than 5 GeV
by one of the straight lines with the
slope we determined in the previous section.

Another model-independent observation we should add about our sample of GRB130427A data
is specific to the particularly significant 9 photons with, energy in excess of 15 GeV.
It is evident in Fig.3 that there are two groups of such photons: a group of 5 photons between 7 and 129 seconds
after trigger, which includes a 72 GeV photon, and a group of 3 photons between 231 and 255 seconds
after trigger, which includes a 95 GeV photon. We should stress, since it will soon be again relevant, that the
arrival times of the last photon of
the first group and of the first photon
of the second group are separated by 102 seconds, which is significantly larger than the separation
between subsequent detection times within each of the two groups.
If we remove ourselves momentarily from the focus on Eq.(1), these qualitative features then should
inspire our characterization of ``hard minibursts".
And notice that
both groups (both hard minibursts)
start with the photon of highest energy in the group and end with the photon of lowest energy
in the group.


We are assuming that looking at GRB130427A is like looking at GRBs in high definition. This allowed us to notice these
features, but we shall now try to use this guidance to go back to previous bright high-energy GRBs
and try to recognize the same features.
As we do this we shall also monitor any indication in favour of redshift dependence of the
features governed by the $D(z)$ of (\ref{main}), which would be the only way for building a case
for a description of these features in terms of quantum-spacetime properties.

\subsection{The case of GRB080916C}
GRB080916C is our first attempt to recognize in another GRB the features here uncovered for GRB130427A.
This is a very interesting GRB for our purposes, particularly for the issue
of possible redshift dependence, since GRB080916C was~\cite{nemiro}
at reshift of 4.05, much higher than the 0.34 of GRB130427A.
And GRB080916C must be viewed as a very powerful high-energy GRB, if one indeed reassesses its LAT fluence
in light of its high redshift.

There is at least one more criterion which would characterize GRB080916C as a GRB
in the Fermi-LAT catalogue whose comparison to GRB130427A is particularly significant
from the viewpoint of the analysis we are here reporting:
GRB080916C contains a 13.3 GeV photon, a 7 GeV photon and a 6.2 GeV photon, and when converting these
   energies at time of detection into energies at time of emission,
   taking into account the large redshift of 4.05,
   one concludes that all these 3 photons had energy at time of
   emission of more than 30 GeV.
This compares well with one aspect which we stressed about GRB130427A, which has 4 photons above 30 GeV
(actually above 40 GeV).

\begin{figure}[htbp!]
\includegraphics[width= 0.89\columnwidth]{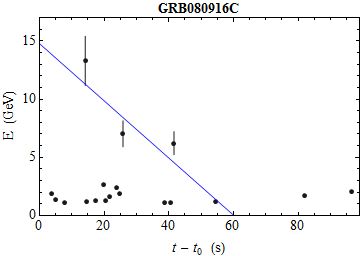}
\caption{Here shown is the Fermi-LAT sequence of photons with energy
greater than 1 GeV for GRB080916C. We visualize error bars at the $15\%$ level
 for the energies of the most energetic events, which play a special role from our perspective.
  The slope of the straight line here shown is not fit on this data but rather is obtained
  from the slope fit in Fig.4 (for GRB130427A) by rigidly rescaling it by the factor of $D(0.34)/D(4.05)$
  dictated by our Eq.(1).}
  \label{fig7}
\end{figure}

Of course, noticing that a GRB contains  photons of such high inferred energies at emission time does not mean we are
assured to see one of our ``hard minibursts with spectral lags". On the contrary the probability of
such success remains rather low, since surely often when we see 3 or 4 such photons they belong to different
hard minibursts within the GRB. To see how this is the case imagine that from GRB130427A we only observed 3 rather than 9
photons with energy greater than 19 GeV: if we take off randomly 6 photons among the top-9 most energetic ones
from Fig.3 chances are no hard miniburst would be noticeable. But in some cases we are bound to get lucky,
and this is what at least appears to be the case with GRB080916C, as shown in Fig.\ref{fig7},
which reports the events for GRB080916C with energy higher than 1 GeV.

Our readers should easily notice from Fig.\ref{fig7} that some features we exposed earlier in this manuscript
for GRB130427A are present also in this GRB080916C case.
In particular, the highest energy photon (13.3 GeV) is detected as first photon after
the longest time interval of ``high-energy silence" during the first 25 seconds.
And  then some other high-energy photons are detected soon after this highest-energy photon.
So actually GRB080916C provides us with a very clear example
of the feature we are proposing, describable as early hard development
of minibursts within a long burst.

Next we should test on GRB080916C the success of (\ref{main}), and look for possible redshift dependence.
For testing redshift dependence governed by the $D(z)$ of (\ref{main})
one should do as in Fig.\ref{fig7}, {\it i.e.} rescale the slope taking into account the
difference of redshifts. According to  (\ref{main}) the relationship between the GRB080916C slope and
the  GRB130427A slope is fixed to  be $D(0.34)/D(4.05)$. The slope
of straight lines characteristic of high-energy spectral lags should be then for GRB080916C much lower
than for GRB130427A.
A straight line with such a slope is shown in Fig.\ref{fig7} and it gives a rather intriguing result:
that straight line obtained from the ones of Sec.III by correcting for the redshift, assuming the validity of (\ref{main}),
is consistent with as many as 3 other GRB080916C events. And one should notice that among the 4 events
that would reasonably fit this straight line one finds the 3 highest-energy photons which we highlighted
at the beginning of this subsection!

\begin{figure}[htbp!]
\includegraphics[width= 0.89\columnwidth]{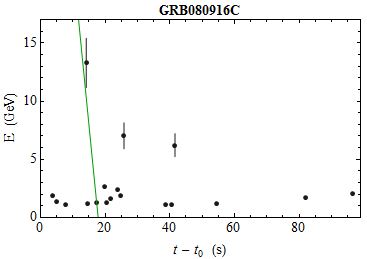}
\caption{Here shown are the same data points already shown in Fig.\ref{fig7}, but now the straight line added on top of the data point has exactly the slope we derived in Sec.III (Fig.4), without any rescaling.}
\label{fig8}
\end{figure}

Of course, the hypothesis that the slope of spectral lags within hard minibursts be governed by Eq.(1) is only
one of many plausible hypotheses. Among these possible alternatives particularly simple is the one
assuming redshift and context independence of the slope of spectral lags within hard minibursts.
This simple hypothesis is probably unrealistic, but at least it gives us an hypothesis alternative
to the redshift dependence of Eq.(1) and this will help us assess the significance of what we find
assuming the redshift dependence of Eq.(1).
In Fig.\ref{fig8} we let go through the point (with error bar) corresponding to the highest-energy
photon a straight line with the slope we determined in the previous section.

Fig.\ref{fig8} shows that by not rescaling the slope for redshift one establishes
no connection between the highest-energy photon and with the second and third most energetic photons in the sample.
This is disappointing for the redshift-independence hypothesis. But of course it may well be that
the remarkable feature highlighted in Fig.\ref{fig7}, for the Eq.(1)-governed-redshift case, is merely accidental
and instead a true connection is the one visible in this Fig.\ref{fig8} for the redshift-independence hypothesis,
with a linear soft spectral lag between the highest-energy photon and the photon detected immediately after it.

But if one relies on just the comparison of  Fig.\ref{fig7} and Fig.\ref{fig8} the conclusion must be
that GRB080916C scores a point for the redshift dependence governed by the $D(z)$ of (\ref{main}),
even though the redshift-independence hypothesis does not fair too badly on GRB080916C data.
Actually the fact that such different
slopes for our linear law of spectral lags within minibursts do not perform in correspondingly widely different
manner in describing some data is due both to the unexceptional quality of the data and
to the main feature we are uncovering: the early hard development
of minibursts is such that we find a sequence with first no high-energy photons for a rather sizable
time interval and then the highest-energy photon followed temporally by several other high-energy photons.
It is evident that with such a structure different slopes for the soft spectral lags
can provide a reasonably good match to the data from the perspective we are here adopting.

\subsection{The case of GRB090926A}
GRB090926A is another long burst of the pre-GRB130427A Fermi-LAT catalogue
that compares well in terms of
the brightness
of the high-energy component with GRB080916.
And GRB090926A was at a redshift \cite{nemiro} of 2.106, and therefore provides a case with
  a value
of redshift somewhere in between the ones for GRB130427A and GRB080916C.

\begin{figure}[htbp!]
\includegraphics[width= 0.71\columnwidth]{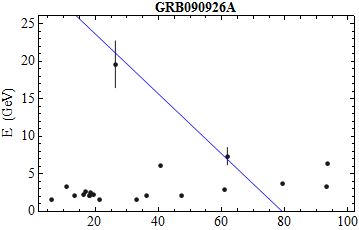}
\includegraphics[width= 0.71\columnwidth]{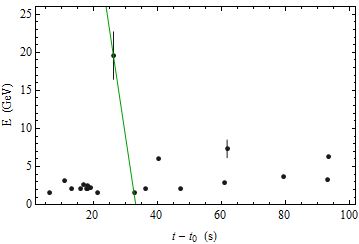}
\caption{Here shown is the Fermi-LAT sequence
of photons with energy
greater than 1.5 GeV for GRB090926A. We visualize an error bar at the $15\%$ level
 for the energy of the two most energetic events.
 The bottom panel also shows a straight line with the slope we derived in Sec.III.
 The top panel also shows a straight line with slope obtained
  from the slope derived in Sec.III by rigidly rescaling it by the factor of $D(0.34)/D(2.106)$
  (2.106 being the redshift of GRB090926A)
  dictated by our Eq.(1).}
\label{fig9}
\end{figure}

In considering GRB090926A we restrict our focus on the photons shown
in Fig.\ref{fig9}, with energy greater than 1.5 GeV.
Again in Fig.\ref{fig9} we see that for GRB090926A (just as observed above for GRB130427A and GRB080916C)
 there is a clearly noticeable ``hard miniburst"
 with early hard development whose onset coincides with the highest-energy photon in the whole signal:
 the figure shows several photons rather densely distributed between 6 and 21 seconds after trigger,
 then ``hard-photon silence" for 5 seconds until at 26 seconds after trigger the highest-energy
 photon in the burst is observed, followed by other high-energy photons.

Concerning the other feature of our interest, involving the presence of soft spectral lags which are linear in energy
  and the possible redshift dependence of this linear law,
  in the spirit of what we did in the preceding subsection we show
  in Fig.\ref{fig10} straight lines on which the highest-energy photon lies (within errors).
  In the bottom panel of Fig.\ref{fig9} the line has exactly the same slope derived in Sec.III
(our redshift-independent scenario) and it is notable that
  the first photon in our sample after the highest-energy photon lies on that line.
  But it is the top panel of Fig.\ref{fig9} which apparently contains the most intriguing message:
  both in GRB130427A and in GRB080916C the slope predicted by Eq.(1) for the same value
  of $M_{QG}$ (and taking into account the redshift) gave a straight line compatible with
  the highest-energy photon and two other of the most energetic photons in the bursts,
  and here again, as shown indeed in Fig.\ref{fig9}, the same strategy of analysis places
  on the same spectral-lag straight line the highest-energy photon and the second-highest-energy
  photon of GRB090926A. Again the slope of the straight  line shown in the top panel of Fig.\ref{fig9}
  was obtained by rigidly rescaling the GRB130427A slope found in Fig.4 by the
  factor $D(0.34)/D(2.106)$ as dictated by the redshift dependence of Eq.(1).

\subsection{The case of GRB090902B}
GRB090902B, which was at a redshift \cite{nemiro} of 1.822,
is the other long burst in the pre-GRB130427A Fermi-LAT catalogue
that compares well in terms of
the brightness
of the high-energy component with GRB080916 and GRB090926A.

\begin{figure}[htbp!]
\includegraphics[width= 0.71\columnwidth]{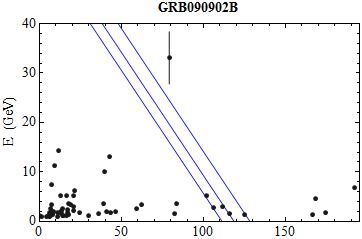}
\includegraphics[width= 0.71\columnwidth]{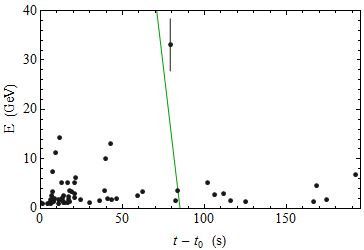}
\caption{Here shown is the Fermi-LAT sequence
of photons with energy
greater than 1 GeV for GRB090902B. We visualize an error bar at the $15\%$ level
 for the energy of the most energetic event.
 The bottom panel also shows a straight line with the slope we derived in Sec.III.
 The top panel also shows some straight lines with slope obtained
  from the slope derived in Sec.III by rigidly rescaling it by the factor of $D(0.34)/D(1.822)$
  dictated by our Eq.(1) for a GRB at redshift of 1.822.}
  \label{fig10}
\end{figure}

   As stressed already in the previous subsection, even in  presence of a few high-energy photons
   one has  no guarantee to see in the data
   a single hard miniburst containing more than one of the highest-energy photons. More often than not the few
   highest-energy  photons Fermi does
   catch  will
   have originated from different hard minibursts within the burst.
   In light of this it may appear that we were unreasonably lucky if Fig.\ref{fig7} for GRB080916C and in the top panel
   of Fig.\ref{fig9} for GRB090926A: our spectral-lag straight lines with slope depending on redshift according
   to Eq.(1) caught the top 3 highest-energy photons of GRB080916C and the top 2
   highest-energy photons of GRB090926A.
   This unreasonably high level of success of our ansatz does not apply to the case of GRB090902B
   which we are now
   considering:
 as we show in the top panel of Fig.\ref{fig10}, focusing on GRB090902B photons with energy greater than 1 GeV,
  our spectral-lag straight lines, with slope computed from the one derived in Sec.III according to the redshift dependence
   coded in Eq.(1) (the correction factor for the slope in this case being $D(0.34)/D(1.822)$),
   do not connect the highest-energy
 photon with any other one of the most energetic photons in GRB090902B.
The top panel of Fig.\ref{fig10} does show that the highest-energy photon of GRB090902B is reasonably compatible
 within our {\it ansatz} with a group of five photons with energy $1 < E \lesssim 5 GeV$ detected between 100 and 125 seconds
 after trigger, but, while this is nonetheless noteworthy, such compatibilities are far less spectacular than the ones
 we found for GRB130427A, GRB080916C and GRB090926A: several photons in GRB090902B have energy $1 < E \lesssim 5 GeV$
 so the (rather weak) compatibility exposed in the top panel of Fig.\ref{fig10} lends itself to being interpreted as accidental,
whereas for GRB130427A, GRB080916C and GRB090926A we always found that our {\it ansatz} established a compatibility
 between at least two (in one case all three) of the top three highest-energy photons.

It is noteworthy that, as shown in the bottom panel of Fig.\ref{fig10}, even in the case of GRB090902B the
redshift-independence hypothesis does not provide a better match to the data than
the hypotesis of redshift dependence governed by Eq.(1).

And even more noteworthy is the fact that even in the case of GRB090902B we cam notice again
the structure of our ``hard minibursts": the sequence of high-energy photons  has a relatively
long silent time interval between the highest-energy
photon and the  photon that precedes it (long compared to  the typical time interval between subsequent
detections of these GRB090902B high-energy photons). So the highest-energy photon is detected after
a sizable time interval of high-energy silence and then a few
high-energy photons are detected soon after.

\subsection{The case of a short burst: GRB090510}
Our analysis of the four Fermi-LAT bursts with most significant high-energy signal, GRB130427A, GRB080916C,
GRB090902B and GRB090926A, gave rather consistent indications. At least one mini-burst with early hard
development was found in all of them in correspondence of the highest-energy photon in the GRB.
And in all four cases the onset of this miniburst containing the highest-energy photon
was preceded by a relatively large time interval of high-energy silence:
one finds in all four these GRBs a suitable lower cutoff on the energies of the photons to be included
in the analysis such that there are no detections of photons in such a sample for a certain time interval
before the detection of the highest-energy photons, while going at still earlier times one finds a denser
rate of detections of photons in the sample.

Concerning the quantification of the soft spectral lags within such minibursts in terms of a linear law of
dependence on energy  GRB080916C,
GRB090902B and GRB090926A, did not add any particularly decisive evidence but offered an overall consistent picture.
As stressed above (particularly for GRB080916C and GRB090926A)
this additional supporting  evidence for the linear quantification
of the soft spectral lags, appears to favour  the hypothesis of redshift dependence governed by Eq.(1),
at least in comparison with the only other hypothesis which for simplicity we are considering,
which is the one of redshift-independent and context-independent slope for the linear law governing
the soft spectral lags.

This being the state of affairs as far as information from long bursts is concerned, evidently
a crucial role could in principle be played by short bursts with strong high-energy signal.
This could for example help discriminating between the interpretation of the features
as properties of the source or as properties of propagation in quantum spacetime: a spacetime-propagation effect
should be present identically for short and long bursts, whereas short and long bursts are sufficiently
different that, in case the features are properties of the sources, going from considering
long bursts to considering short burst might make a big difference.

\begin{figure}[htbp!]
\includegraphics[width= 0.71\columnwidth]{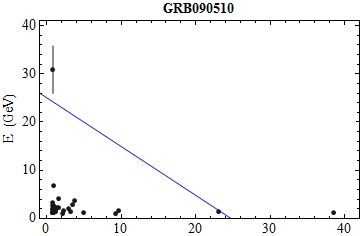}
\includegraphics[width= 0.71\columnwidth]{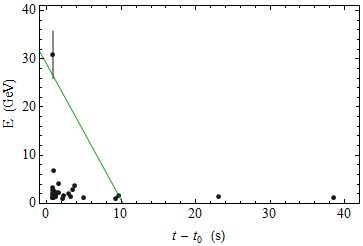}
\caption{Here shown is the Fermi-LAT sequence
of photons with energy
greater than 1 GeV for GRB090510. We visualize an error bar at the $15\%$ level
 for the energy of the most energetic event.
 The bottom panel also shows a straight line with the slope we derived in Sec.III.
 The top panel also shows a straight line with slope obtained
  from the slope derived in Sec.III by rigidly rescaling it by the factor of $D(0.34)/D(0.897)$
  (0.897 being the redshift of GRB090510), as
  dictated by our Eq.(1).}
  \label{fig11}
\end{figure}

There is a single short burst in the Fermi-LAT catalogue with a sufficiently strong signal above 1 GeV
for allowing the sort of analysis we are performing. This is GRB090510, whose sequence of detection times
for photons with energy greater than 1 GeV is shown in Fig.\ref{fig11}.

In this case let us start by discussing our test with the spectral-lag straight lines
derived in Section III from GRB130427A data.
As shown in Fig.\ref{fig11} one does find some potential candidates as high-energy photons lagging
the highest energy photon according to the quantification we first derived
in Section III, but with respect to the outcome of the analogous tests we did previously on long bursts
in this case the outlook is less promising: both for the redshift-independent hypothesis and for the Eq.(1)-based
redshift-dependent
hypothesis the test does not fail badly but it also gives only marginally plausible candidates.
This is because in Fig.\ref{fig11} one notices that the highest-energy photon lies in time-of-detection quasi-coincidence with
several other high-energy photons. It is hard to believe that this time coincidence is accidental
(not a manifestation of an inter-relation between the highest-energy photons an those other
nearly coincident high-energy photons). Both the top and bottom panel of Fig.\ref{fig11} would suggest
that the highest-energy photon happens to have been detected in reasonably good time coincidence with those other
high-energy photons but actually it is part of a miniburst involving photons detected at much later times.
This is certain possible, but we feel it is somewhat hard to believe.

 The fact that in the top panel and even better in the bottom panel of Fig.\ref{fig11} we did find
at least some candidates for
 the type of correlations predicted by our linear law of soft spectral lags leaves open the door
 for hoping that our linear law of soft spectral lags might be applicable also to short GRBs.
 Moreover, we notice that in the Fermi-LAT data on GRB090510 there is only one photon with energy greater
 than 30 GeV (indeed the 31-GeV event we discussed), and, considering that  the redshift of GRB090510 was\cite{nemiro} 0.897,
there was no other photon in GRB090510 data set with energy greater than 30 GeV
even at time of emission. If a criterion such as a cutoff at 30 GeV in energy at emission was
 to govern the features we are here exposing, then GRB090510 should not even be studied from our
 perspective.

\begin{figure}[htbp!]
\includegraphics[width= 0.7\columnwidth]{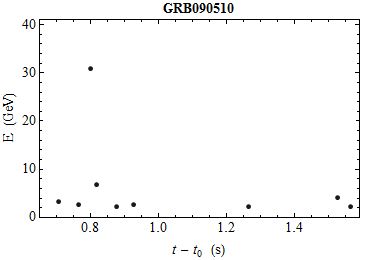}
\caption{A subset of the GRB090510 data points already shown in Figs.\ref{fig10} and \ref{fig11}: the ones here included have energy
  in excess of 2 GeV and where detected between 0.7 and 1.6 seconds after the trigger of GRB090510.}
\label{fig12}
\end{figure}

This noticed about the spectral-lag story for GRB090510, we still need to explore the implications
of GRB090510 for the other side of the features we are exposing, which concerns the presence and characteristics
of a hard miniburst with onset marked by the highest-energy photon in the GRB.
The fact that the highest-energy photon lies in time-of-detection quasi-coincidence with
several other high-energy photons renders our description of hard minibursts apparently inapplicable to this short burst,
at least on the sort of time scales at which instead this feature was systematically present among the long bursts
we analyzed.
We then looked for the possible presence of this feature on shorter time scales and including
in the description only photons with energy greater than 2 GeV. Fig.\ref{fig12} shows a snapshot of the content of
the previous Fig.\ref{fig11}, restricted to times between 0.7 and 1.6 seconds after trigger and to photons with energy
greater than 2 GeV. And in Fig.\ref{fig12} one could tentatively recognize one of our hard minibursts with onset marked by the
highest-energy photon: to the left of the highest-energy photon Fig.\ref{fig12} shows one photon 0.025 seconds before it
and another photon 0.099 seconds before it, while to the right of the highest-energy photon with 31-GeV energy first comes
a photon with energy of 6.7 GeV just 0.017 seconds after it and another photon with energy of 2.3 GeV some 0.076 seconds after it.

\subsection{Off the record}
In this subsection we offer a side the remark, which readers are asked to perceive as off the record, intended
only as a small motivator for future searches on another feature which may arise as an unnecessary but plausible
corollary manifestation of the high-energy spectral lags we are noticing.
This is about the possibility that some high-energy photons be detected before the GRB trigger. None has been reported
so far in studies of Fermi-LAT GRBs, so evidently the data must not provide safe candidates for such early detections.
We nonetheless looked for any multi-GeV photons detected by LAT within 100 seconds before the five GRBs here considered.
Also for this search we adopted the criteria and methodology described in Subsection II.A, by retrieving
LAT Extended Data, processing these data with the Fermi Science Tools,
and using as selection criterion the ``P7TRANSIENT'' class or
better.

We found two candidates worth mentioning (off the record): a 80 GeV photon some 25 seconds before the trigger
of GRB090926A and a 16 GeV photon some 7 seconds before the trigger of GRB090510. Both of these are transient events,
with a correspondingly tangible risk of not being truly connected to the relevant GRBs.
The 80 GeV photon is also found
at about  6 degrees from the optical position of GRB090926A, which is still plausible for
a genuine GRB photon, but is an angle large enough to further weaken the case of its association with  GRB090926A.
The status of the 16 GeV candidate is somewhat more robust in this respect, since it is found within just 3.8 degrees
of GRB090510.

Since we are going off the record we shall not dwell any further on the robustness of these two events.
We rather prefer to show how they would change the outlook of our analysis of GRB090926a (Subsec.IV.C)
and of GRB090510 (Subsec.IV.D).

\begin{figure}[htbp!]
\includegraphics[width= 0.71\columnwidth]{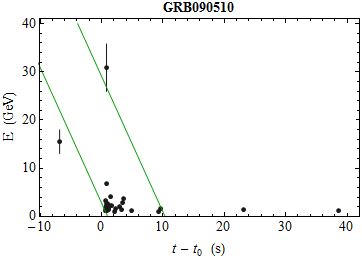}
\includegraphics[width= 0.71\columnwidth]{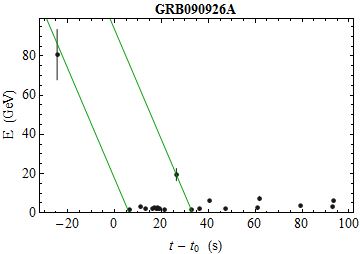}
\caption{Here shown in the top panel is the same data and slope of straight lines already shown in the bottom panel
of Fig.\ref{fig11}, for GRB090510, but now adding a precursor candidate some 7 seconds before trigger.
The bottom  panel here shows the same data and slope of straight lines already shown in the bottom panel
of Fig.\ref{fig10}, for GRB090926A, but now adding a precursor candidate some 25 seconds before trigger.}
\label{fig13}
\end{figure}

In Fig.\ref{fig13} we do this for the hypothesis that the spectral-lag slope we found analyzing GRB130427A data
should be applied to all bursts, independently of redshift. It is striking that both of the candidate percursor multi-GeV
events would be placed in reasonably good time correlation with the trigger of the GRB by this spectral-lag slope.

\begin{figure}[htbp!]
\includegraphics[width= 0.71\columnwidth]{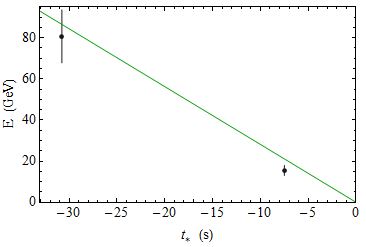}
\caption{The energies of our candidate precursors of 80 GeV and 16 GeV  (repspectively to GRB090926A and GRB090510) are
here shown in correspondence of time values corresponding the the time interval by which each of them preceded
the first GeV event of corresponding GRB. The slope of the straight line here shown is the one derived in Sec.III.}
\label{fig14}
\end{figure}

In Fig.\ref{fig14} we show how this could be arranged in the form of a general behaviour for such candidate percursor multi-GeV
events. Fig.\ref{fig14} just plots the energy of our two candidates as function of the time advancement found with respect to
the first ``official" (``on record") GeV photon
photon with energy greater than a GeV in the GRB which might be associated to them,
and we superpose again a straight line with the slope determined
in Section III and this time assuming that it should cross the origin (reflecting the assumption that our linear spectral lag
should describe exactly the time advancement with respect to the first GeV event on record).

The content of Fig.\ref{fig14} is rather impressive, but with only two candidate events (one of which is a particularly weak
candidate) the whole content of Fig.\ref{fig14} may well be attributed to chance.
But although we consider this all subsection ``off the record" we do feel that one of the massages implicitly
contained in Fig.\ref{fig14} should be taken into account: rare multi-GeV candidate precursors have been so far dismissed
completely because they happened to be
events that could only be unreliably associated to a corresponding GRB,
but it must be appreciated that as a larger statistics builds up of such weak candidate multi-GeV precursors
the collective picture they provide might be reliable even if each individual member of the family
is not fully reliable. Imagine for example that we reach at some point the situation where, say, 50 such
candidate multi-GeV precursors have been accumulated and they all (or nearly all) happened to line up
on the straight line in Fig.\ref{fig14}. That hypothetical  collection of candidate precursors would provide robust
support for the straight line in Fig.\ref{fig14} even if each of them individually could not be viewed as a robust candidate.

\begin{figure}[htbp!]
\includegraphics[width= 0.71\columnwidth]{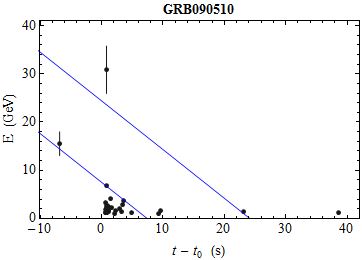}
\includegraphics[width= 0.71\columnwidth]{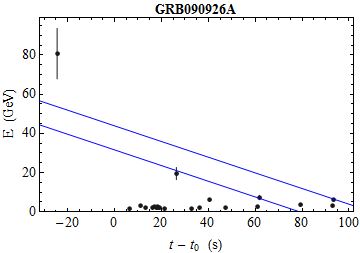}
\caption{Here shown in the top panel is the same data and slope of straight lines already shown in the top panel
of Fig.\ref{fig11}, for GRB090510, but now adding a precursor candidate some 7 seconds before trigger.
The bottom  panel here shows the same data and slope of straight lines already shown in the top panel
of Fig.\ref{fig10}, for GRB090926A, but now adding a precursor candidate some 25 seconds before trigger.}
\label{fig15}
\end{figure}

In closing this peculiar subsection we look at our two candidate multi-GeV precursor events from the perspective
of the redshift dependence codified by the quantum-spacetime picture of Eq.(\ref{main}).
This is done in Fig.\ref{fig15}.

The top and bottom panels of Fig.\ref{fig15} show the same data points
in the top and bottom panels of Fig.\ref{fig13} but now the superposed straight lines have slope obtained by taking
the slope derived in Section III for GRB130427A and rescaling that slope by taking into account the differences
in redshift precisely as codified in Eq.(\ref{main}). Also Fig.\ref{fig15} potentially contains powerful messages. In particular
the 16 GeV candidate precursor of GRB090510 is consistent with one such straight line (top panel) which includes two other photons,
one of which is the one with the second highest energy among those ``on record" for GRB090510.
And a similar message is contained in the bottom panel of Fig.\ref{fig15}, though with less significance: the 80 GeV candidate
precursor of GRB090926A is consistent with a straight line governed by (\ref{main}) which also includes
the photon with the second highest energy among those ``on record" for GRB090510.

\subsection{A speculation about GRB neutrinos}
In the same spirit of the considerations offered in the previous subsection we actually can also consider
some candidate GRB neutrinos. This however is already ``on record" since we will basically
look from our perspective at the analysis announced in Ref.\cite{gacDAFNEpiran} (form March 2013,
safely {\underline{before}} the observation of GRB130427A).

Ref.\cite{gacDAFNEpiran} also contemplates the type of energy dependence of times of arrival
that is produced by Eq.(\ref{main}), exactly as we are doing in the present study.
But Ref.\cite{gacDAFNEpiran} was aimed at the context of GRB-neutrino searches by IceCube.

Recently IceCube reported \cite{icecubenature} no detection of any  GRB-associated neutrino in a data set taken from April 2008 to May 2010. Ref.\cite{gacDAFNEpiran} focuses however on the 3 Icecube neutrinos that one can find in published material \cite{icecubenature,icecubetesi}
and that are plausible (though weak) GRB-neutrino candidates.
These are \cite{icecubetesi}
a 1.3 TeV neutrino {  1.95$^o$ off GRB090417B, with localization uncertainty of 1.61$^o$, and detection time  2249 seconds {\underline{before}} the  trigger of GRB090417B,
a 3.3 TeV neutrino  6.11$^o$ off GRB090219, with a localization uncertainty of 6.12$^o$, and detection time  3594 seconds {\underline{before}} the GRB090219 trigger and  a 109 TeV neutrino, within 0.2$^o$  of GRB091230A, with a localization uncertainty of 0.2$^o$, and detected some 14 hours {\underline{before}} the GRB091230A trigger.

The fact that all  3 of these GRB-neutrino candidates
were detected sizably in advance of the triggers of the GRBs they could be associated with
is not particularly significant from the standard perspective of this sort of analysis,
and actually obstructs any such attempt to view them as  GRB neutrinos:
no current GRB model suggests that neutrinos could be emitted thousands of seconds before a GRB.
But a collection of GRB-neutrino candidates all sizably in advance of
corresponding GRB triggers is just what one would expect from (\ref{main}) (for $s_\pm =1$).
Using this observation in Ref.\cite{gacDAFNEpiran} it was shown that possibly 2 of the 3 mentioned
GRB-neutrino candidates could be actual GRB neutrinos, governed by  (\ref{main}).

The most appealing possibility discussed in  Ref.\cite{gacDAFNEpiran} is for the 3.3 TeV and the 109 TeV neutrinos to be actual GRB neutrinos,
and requires $M_{QG} \simeq M_{planck}/25$.
The other possibility is for the 1.3 TeV and the 3.3 TeV neutrinos to be actual GRB neutrinos,
and requires an even smaller value of $M_{QG}$, of about $M_{QG} \simeq M_{planck}/100$.

Evidently what we found in the present analysis inspired and based in part on GRB130427A
is at least to some extent
consistent with what was found in
Ref.\cite{gacDAFNEpiran} for the 3.3 TeV and the 109 TeV neutrinos. In order to be more precise
in this assessment we must take into account the fact that the redshifts of GRB090219 and of GRB091230A
were not determined.
We shall deal with this in the next (closing) section by handling the uncertainty in the redshift as another source of
uncertainty for testing  (\ref{main}) with these two neutrinos:
since  GRB090219 was a short burst we can reasonably \cite{wandeLONG} assume its redshift was between 0.2 and 1,
whereas for GRB091230A, a long burst, we can reasonably \cite{cowaSHORT} assume its redshift was between 0.3 and 6.

\newpage

\begin{figure*}[htbp!]
\includegraphics[width=1.65 \columnwidth]{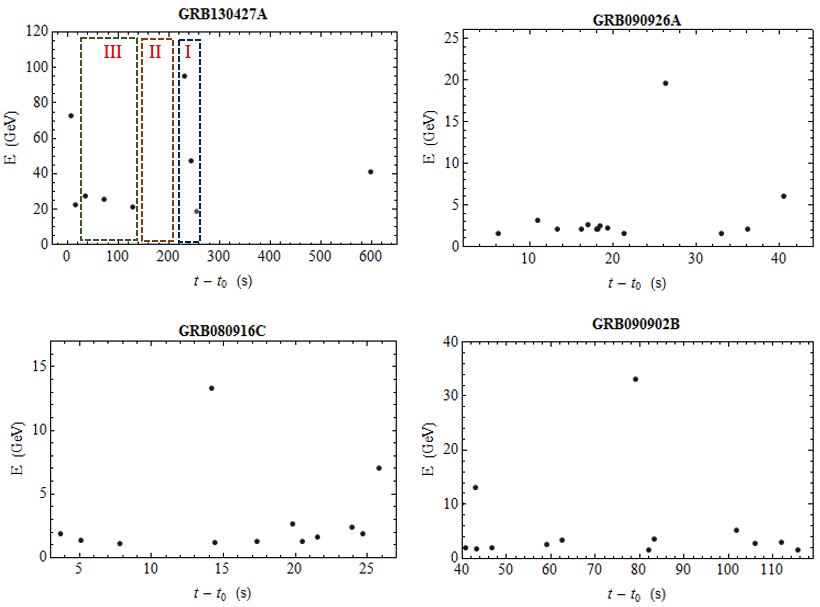}
\caption{In the top-left panel we here use GRB130427A data points with energy greater than 15 GeV
for illustrating the structure of
our hard minibursts. The actual miniburst (I) is preceded by a relatively long time interval of ``high-energy silence" (II)
and farther back in time one finds rather frequent detections of high-energy photons (III).
The first photon in the hard miniburst is the one with highest energy. The other panels here shown summarize the supporting
evidence for this structure of hard minibursts which we found analyzing the three brightest pre-GRB130427A Fermi-LAT long GRBs.}
\label{fig16}
\end{figure*}

\section{Perspective on results and closing remarks}
We here took as starting point the remarkable Fermi-LAT observation of GRB130427A.
We looked at this GRB assuming it was giving us a chance to see a long bright burst in high resolution.
Using as guidance the quantum-spacetime-inspired {\it ansatz} of Eq.(\ref{main}) we stumbled upon
a few related features present in the highest-energy portion of the Fermi-LAT data GRB130427A.
We then used this insight gained from the high-resolution GRB130427A to recover
some evidence of the same features in the brightest previous Fermi-LAT observations of GRBs, and even
for some candidate GRB neutrinos.

Since we considered a few features in several contexts it is useful to close this manuscript by
offering a perspective on our findings.
We articulate this section into subsections, each one devoted to one of the features and aspects we have considered,
and ordered according to the strength of the supporting evidence which we found.

\subsection{Structure of hard minibursts}
The feature that emerged most robustly from our analysis concerns the presence of hard minibursts, whose onset is marked
by the highest-energy photon observed in a given GRB. We found this feature (here illustrated in Fig.\ref{fig16}) in all the top four
most powerful high-energy GRBs in the Fermi-LAT catalogue: GRB130427A, GRB080916C, GRB090902B and GRB090926A.

The minibursts become noticeable only when focusing the analysis on the few highest-energy photons in the signal:
for GRB130427A it was clearly visible for photons with energy higher than 5 GeV, while for GRB080916C, GRB090902B and GRB090926A
it manifested itself  when restricting the analysis to photons of energy higher than 1 GeV or 2 GeV.
The structure of these hard minibursts was in all cases such that
the high-energy signal (composed of photons selected with energy above a certain high value,
as we just stressed) has a long silent time interval between the highest-energy
photon and the high-energy photon that precedes it (long compared to  the typical time interval between subsequent detections
of these high-energy photons). So the highest-energy photon is detected after
a sizable time interval of high-energy silence of the signal, and then a few
high-energy photons are detected soon after.

Besides being identifiable in connection with the highest-energy photon of all the four brightest long Fermi-LAT GRBs,
GRB130427A, GRB080916C, GRB090902B and GRB090926A, in the case of GRB130427A one can tentatively identify
two other such hard minibursts, with exactly the same structure.

Only one short burst in the Fermi-LAT catalogue is suitable for this sort of analysis, so for short bursts
our claims about hard minibursts are inconclusive, but the content of Fig.\ref{fig12} may hint that such hard minibursts
are also present in short GRBs.

\subsection{Soft spectral lags within hard minibursts of GRB130427A}
The structure we exposed for the hard minibursts evidently produces some soft spectral lags,
at least to the extent that we found in all cases the onset of the miniburst was marked by the
highest-energy photon.

But we have some evidence for a specific behaviour of these soft spectral lags. Evidence of such soft spectral lags was
of course clearest in the case of GRB130427 for which we managed to deduce a behavior
of the spectral lags going linearly with energy. This was noticed in Subsection III.A by exhibiting some rather
compelling candidate photons for 3 hard minibursts, and was shown in  Subsection III.B by a statistical
analysis taking the shape of a correlation
study determining the frequency of occurrence of spectral lags in our GRB130427A sample.

GRB130427A on its own provides a rather compelling case for the presence of these linear-in-energy soft
spectral lags. And when we considered available data on other bright Fermi-LAT GRBs we did find some additional
supporting evidence, though tangibly weaker than in the GRB130427A case.

\subsection{Strength and weakness of the quantum-spacetime-inspired hypothesis}
When we looked to other bright Fermi-LAT GRBs as opportunities for  additional
 evidence supporting our linear spectral lags
 it was necessary to contemplate a redshift dependence
of the linear law. For no better reason than familiarity, we had a candidate ready for the task,
taking the shape of the redshift dependence codified by $D(z)$ for the linear spectral lags of (\ref{main}).
We went on to explore the efficaciousness of this hypothesis in reproducing features of GRB data.
This means that we took the  linear energy-vs-time dependence derived on GRB130427A data and we assumed that
it should apply to other bursts by rigidly rescaling the slope by $D(0.34)/D(z)$,
where $D(0.34)$ is the $D(z)$ of GRB130427A (with redshift indeed of 0.34)
and the $z$ in $D(z)$ is the redshift of the other burst to which one intends to apply
the results obtained for GRB130427A.

A dramatic confirmation would have been to find for each burst several photons lining
up (in energy vs time of detection) just according to the linear relationship, with slope
rigidly rescaled by $D(0.34)/D(z)$. But of course this was impossible since even for the brightest pre-GRB130427A Fermi-LAT
long bursts, GRB080916C, GRB090902B and GRB090926A the LAT only detected a rather small overall number
of photons with energy greater than, say, 5 GeV.
Nonetheless at least in two cases our investigation of this possibility
appears to have been successful even beyond what one (at least we) could
reasonably expect.
For GRB080916C by rigidly rescaling the slope of our linear law according to $D(0.34)/D(z)$ (no freedom
to tweak there) and insisting that the relevant straight line be consistent with
the highest-energy photon (introducing only freedom governed by the uncertainty in the energy of the highest photon,
which we assumed to be $\simeq 15\%$) we found a rather noteworthy group of four photons
with energy and detection times all consistent with our linear law for soft spectral lags.
These are the four photons on the straight line in  Fig.\ref{fig7}, which include in particular the highest-energy photon
and the second and third most energetic photons of GRB080916C.
And in the analogous analysis of GRB090926A we found that our {\it quantum-spacetime ansatz} for the dependence on
redshift of the linear spectral lags remarkably also placed in connection the two highest energy photons.
While all this may well be accidental, the cases of  GRB080916C and GRB090926A
clearly can seen as proviing encouragement for the law of redshift dependence of Eq.(\ref{main}),
and with it the possible quantum-spacetime interpretation of the features we here exposed.
We must stress that if we could  fully establish the dependence on redshift given by $D(z)$ (or one of its very few variants
allowed by the quantum-spacetime interpretation \cite{dsrds})
the spacetime origin of this feature would have to be assumed, with or without
the full content of Eq.(\ref{main}).

Let us therefore summarize the most striking possible elements in support of this spacetime interpretation
which were encountered in our analysis.
For this purpose, since we are going to compare data obtained from GRBs at different
redshifts,
it is useful to look at data from the perspective of the following way to
rewrite Eq.(\ref{main}) (restricting of course our focus
to $s_\pm =1$):
$$E  = \frac{c M_{QG}}{D(0.34)}\tau_{QG}(z)$$
with
$$\tau_{QG}(z)  = - \frac{D(0.34)}{D(z)}t_{QG}$$
By rewriting Eq.(\ref{main}) in this way we can place data from different GRBs (at different redshifts)
on the same $E$-vs-$\tau_{QG}(z)$ plots. The change of perspective we are now adopting is to compare data not by
rescaling the slope on the basis of redshift, but instead equivalently rescaling the measured times  $t_{QG}$
in such a way as to englobe the redshift dependence of ${D(z)}$
 in the new time observable $\tau_{QG}(z)  = - D(0.34) t_{QG}/{D(z)}$.

 This allows us to produce the Fig.\ref{fig17}.

\begin{figure}[htbp!]
\includegraphics[width=0.95 \columnwidth]{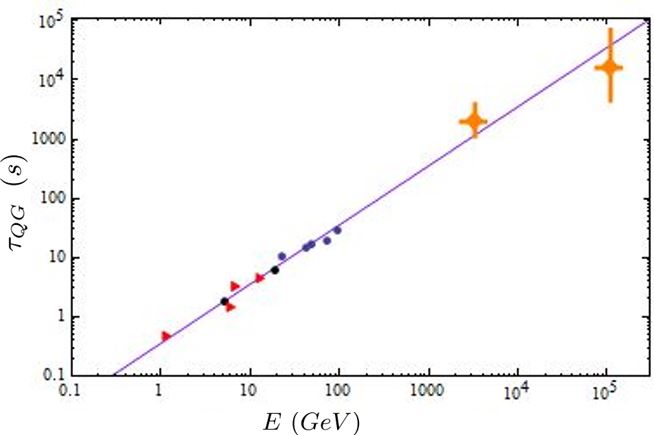}
\caption{We here show in a loglog plot a straight line with the slope we determined in Sec.III.
Also shown (as thick points) are the 7 data points for GRB130427A already shown in Fig.4. The red triangles
describe data on four events found on a single spectral-lag straight line in Fig.\ref{fig7} for GRB080916C.
The orange point with sizable error bars are for the 3.3TeV neutrino and the 109TeV neutrino discussed
from our perspective in Ref.\cite{gacDAFNEpiran}, with 30 percent uncertainty on neutrino
energies reflected in horizontal segments and vertical segments reflecting the
uncertainty in redshift of the GRB possibly associated to the two neutrinos.}
\label{fig17}
\end{figure}

In this Fig.\ref{fig17}  we collect the 7 events from GRB130427A already considered in Fig.4, the mentioned 4 events from GRB080916C
 that are consistent with our Eq.(1)-based straight line of soft spectral lags,
 and we also show the two candidate GRB neutrinos of 3.3 TeV and 109TeV discussed in Subsection IV.F.
 The horizontal orange segments reflect a $30\%$ energy uncertainty for the neutrinos, while the vertical
 segments reflect an uncertainty in the computation of $\tau_{QG}(z)$ for those two neutrinos, which, as mentioned,
 is due
 to the fact that the redshifts of the GRBs to which they are tentatively associated is not known.
Specifically the two neutrino points are for a 3.3 TeV (with $30\%$ uncertainty) neutrino
with $t_{QG}$ of $-3594seconds$ assumed to have originated from a source at redshift between 0.2 and 1
(the short GRB090219)
and for a 109TeV (with $30\%$ uncertainty)
neutrino with $t_{QG}$ of $-14 hours$ assumed to have originated from a source at redshift between 0.3 and 6
(the long GRB091230A).

 Fig.\ref{fig17} contains information from 13 data points,
 much of which was unconstrained by the setup of the analysis. We already discussed in Section III
 how the information contained in the seven points from GRB130427A is only in part used to the
 determine the slope of our linear relation between energy and time of detection and the parameters
 of rigid  time translation needed to align the 3 straight lines from Fig.3. Since the slope of the straight line in Fig.\ref{fig17}
 is still the one determined using that procedure, the information in
 the remaining 6 points in Fig.\ref{fig17} was largely unconstrained within our analysis. The two neutrino points are located
 in the figure just where they naturally should be: for spectral lags of thousands of seconds with respect
 to a GRB trigger there is nothing to be gained by leaving some freedom for the actual
 emission time of the neutrino within the GRB event.
 The information in the 4 points from GRB080916C was nearly all unconstrained: the points where only used to
 determine the  single parameter
 of rigid  time translation needed to reset the onset of that candidate hard miniburst of GRB080916C conveniently
 at $\tau_{QG}=0$.

 Since so much of the information from the 13 points in Fig.\ref{fig17} had been left unconstrained
 and yet the message of Fig.\ref{fig17} is so compelling, clearly the quantum-spacetime interpretation
 of the features we here exposed
 deserves further investigation in future studies and particularly with future observations.

 But in spite of the compelling message contained in Fig.\ref{fig17} we cannot fail to notice some indications
 which instead should recommend prudence in adopting the quantum-spacetime interpretation.
 A first indication of this sort merely originates from theory prejudice: as mentioned
 there are solid models
 of spacetime quantization that predict a behaviour of the type given in Eq.(1) but none of these known models
 predicts that the onset of the validity of Eq.(1) should be abrupt at some energy scale. We get the compelling picture
 of Fig.\ref{fig17} focusing on a few high-energy photons from GRBs, but the possibility that
  the spectral lags predicted by the straight line in Fig.\ref{fig17} apply also to the abundant observations of lower-energy photons
 is already excluded experimentally\cite{fermiNATURE,unoSCIENCE,ellisPLB2009,hessREVIEW,nemiro,gacsmolin}.
 And, as mentioned, the available literature suggests that at energies above the range which was here
 considered for Fermi-LAT GRBs observations one should take into account data from
 observations of some Blazars which also exclude the applicability of Eq.(1) with $M_{QG}$ as low
 as $M_{planck}/25$.

 Of course we could (and eventually would) set aside this theory prejudice in the face of a clear
 experimental situation showing the validity of Eq.(1) only in a rather specific energy range.
 But even then there is an interpretational challenge in the data which mainly resides
 in a single but very well observed photon: the 31 GeV event in GRB090510.
 That 31 GeV event occurred during a time of about 0.2 seconds for which Fermi's LAT
 observed a rather intense burst of photons with energy between 1 and 2 GeV. Our Fig.\ref{fig11} suggests that one could interpret
 that 31 GeV event consistently with  Eq.(1) and consistently with the slope determined in Fig.4 from GRB130427A data
by arguing that the coincidence in timing is accidental: the 31 GeV event happened to occur at
 some point during those special 0.2 seconds but there would be no connection between the mechanism and time of emission
 of the 1-to-2 GeV photons being detected in that 0.2 seconds interval and the mechanism and time of emission
 of that 31 GeV photon. This is not impossible (and Fig.\ref{fig11} itself offers a ``storyline" for accommodating such
 an interpretation) but it is hard to believe.

\subsection{Aside on the redshift-independent hypothesis}
The only other case we here contemplated in alternative to the case of redshift dependence of the slope
governed by the $D(z)$ of Eq.(1) is the case of a slope that is redshift independent and context independent.
For us this served exclusively the purpose of placing in some perspective
the outcome of our investigation of the hypothesis
of redshift dependence governed by  Eq.(1).
Of course if one was to exclude the interpretation as a quantum-spacetime/propagation effect,
it would be most natural not to think of some redshift independent and context independent features,
but rather think of  intrinsic features, of astrophysical origin.

It should also be noticed that in our exploration the redshift-independent hypothesis
provided no noteworthy observations.
And the timing of the 31 GeV photon within GRB090510 appears as unnatural form within the redshift-independent
hypothesis as it does for the quantum-spacetime hypothesis of redshift dependence governed by $D(z)$ of Eq.(1).
Of course, if the features here exposed are intrinsic features one would expect them to depend on the context
of the specific GRB under consideration ({\it e.g.} long vs short), and then the 31 GeV photon within GRB090510
may well find a natural description within the hard minibursts and the soft spectral lags here discussed.

\subsection{Outlook}
The most fascinating interpretation of the features here discussed clearly would
be the one based on quantum-spacetime effects. This quantum-spacetime interpretation requires
 that the effects are present in exactly the same way in all bursts,
with the only differences allowed being governed by redshift, through a function of redshift
of the type of the $D(z)$ of Eq.(1).
One is allowed (in spite of this being something that lacks any support in the quantum-spacetime
 formalisms so far explored) to speculate about possible scales of abrupt onset or offset of the spacetime effects,
 but a dependence on redshift
of the type of the $D(z)$ is non-negotiable for the quantum-spacetime interpretation.
At least on the basis of the
consistency between Fig.3 for GRB130427A and Fig.\ref{fig7} for GRb080916C
further exploration of this interpretation is to some extent encouraged.
But the specific model based on Eq,(1), extending over all values of energy,
 is clearly inadequate for such explorations
because of the limits established for effects of this magnitude
in previous studies of GRB photons at energies below the ones that played a key role here
and in previous studies of blazars at energies above the ones that played a key role here.
Moreover, as we also stressed, the quantum-spacetime interpretation of the features here exposed
is significantly challenged by the applicability to GRB090510, with its
 31 GeV photon detected in sharp time coincidence with several photons of energy between 1 GeV and 2 GeV.

Considering the outmost importance of the scientific issues at stake the quantum-spacetime
avenue should be nonetheless pursued vigorously, but of course it is at least equally natural
to seek a description of the features here highlighted on the basis
of intrinsic properties of the sources, as a pure astrophysics interpretation.
 We made room here for this possibility only to the extent that we considered
 the case of a redshift-independent linear law of soft spectral lags.
 But evidently if the interpretation of the features here uncovered is based on intrinsic properties
 of the sources the quantifications are likely to be context dependent.
An example of avenue for a description based on intrinsic properties of the sources can be inspired
by the fact that the features here discussed all are relevant at very high energies:
this may suggest an inverse-compton origin, which in turn may allow a description of the soft spectral lags
  in terms of cooling time being faster at progressively high energies.

  It is intriguing that
  even if spectral lags within our hard minibursts do not have a quantum-spacetime origin their
  understanding will be important for quantum-spacetime research: the search of possibly minute
  quantum-spacetime-induced spectral lags evidently requires as a prerequisite
  our best possible understanding
  of intrinsic mechanisms at the sources that can produce spectral lags.
  As the understanding of our hard minibursts improves
  they may well turn into ``standard candles", with
  distinctive standard features, ideally suited for the search of possible additional
  contributions to the spectral lags with characteristic dependence on redshift
  of the type $D(z)$ and therefore attributable to quantum-spacetime effects.
  
\newpage

\noindent
{\it Note added: As we were in the final stages of preparation of this manuscript we noticed
on arxiv.org the manuscript arXiv:1305.1261~\cite{cina130427a} which also concerns GRB130427A.
The results obtained processing
the Fermi LAT data on GRB130427A
 here and in arXiv:1305.1261 are perfectly consistent with each other, when taking into account
the somewhat different event-selection criteria used, which reflect the different scientific objectives of
the two studies. Indeed the scientific objectives of the two analyses of the data have no overlap.}

\bigskip
\bigskip
$~$\\$~$\\

\bigskip
\bigskip
$~$\\

This research was supported in part by the John Templeton Foundation (GAC).


\begin{thebibliography}{50}

\bibitem{zhu} S. Zhu, J. Racusin, D. Kocevski et al. 2013 GCN CIRCULAR 14471

\bibitem{kienli} A. von Kienli reports on behalf of the Fermi GBM Team 2013 GCN CIRCULAR 14473

\bibitem{maselli} A. Maselli, A. P. Beardmore, A. Y. Lien et al. 2013 GCN CIRCULAR 14448

\bibitem{golene} S. Golenetskii, R. Aptekar, D. Frederiks et al. 2013 GCN CIRCULAR 14487

\bibitem{pozanenko} A. Pozanenko, P. Minaev, A. Volnova  2013 GCN CIRCULAR 14484

\bibitem{verrecchia} F. Verrecchia, C. Pittori, A. Giuliani  et al. 2013 GCN CIRCULAR 14515

\bibitem{grb881024} A. Chernenko and I. Mitrofanov,
Mon. Not. R. Astronom. Soc. 274 (1995) 361.


\bibitem{flores} H. Flores, S. Covino, D. Xu et al. 2013 GCN CIRCULAR 14491

\bibitem{levan}  A.J. Levan, S.B. Cenko, D.A. Perley et al. 2013 GCN CIRCULAR 14455


\bibitem{xu} D. Xu, C. Cao, S.-M. Hu, et al.  2013 GCN CIRCULAR 14458

\bibitem{elenin} L. Elenin,  A. Volnova,  V. Savanevych et al. 2013  GCN CIRCULAR 14450

\bibitem{acke} M. Ackermann, M. Ajello, L. Baldini,  et al.  ApJ 726 (2011) 81

\bibitem{atwood2009} W. B. Atwood,  A. A. Abdo, M.  Ackermann et al ApJ 697 (2009) 1071

\bibitem{acke2012} M. Ackermann, M. Ajello, A. Albert  et al.  ApJS 203 (2012) 4

\bibitem{grbgac} G.~Amelino-Camelia, J.~Ellis, N.E.~Mavromatos,
D.V.~Nanopoulos and S.~Sarkar,
Nature {393} (1998) 763 [arXiv:astro-ph/9712103]

\bibitem{gampul} R.~Gambini and J.~Pullin,
{Phys.~Rev.}~{D59} (1999) 124021.

\bibitem{urrutia} J.~Alfaro, H.A.~Morales-Tecotl and L.F.~Urrutia,
Phys.~Rev.~Lett.~84 (2000) 2318.

\bibitem{gacmaj} G.~Amelino-Camelia and  S.~Majid,
Int.~J.~Mod.~Phys.~A 15, 4301 (2000) [arXiv:hep-th/9907110]

\bibitem{myePRL} R.C. Myers and M. Pospelov,
    Phys. Rev. Lett. 90, 211601  (2003) [arXiv:hep-ph/0301124]

\bibitem{jacobpiranDELAY}
 U. Jacob and T. Piran,
    JCAP 0801, 031  (2008) [arXiv:0712.2170]

\bibitem{mattiLRR} D. Mattingly, Living Rev.Rel. 8, 5  (2005) [arXiv:gr-qc/0502097]

\bibitem{gacsmolin} G. Amelino-Camelia and L. Smolin, Phys.Rev. D80, 084017 (2009)
 [arXiv:0906.3731]

\bibitem{szabo} R.J. Szabo, Phys.Rept. 378 (2003) 207.

\bibitem{fermiNATURE} A.A. Abdo
{\it et al}.
[Fermi LAT/GBM Collaborations], Nature 462, 331 (2009).

\bibitem{unoSCIENCE}
  A. Abdo et al,
  Science 323 (2009)
  1688

\bibitem{ellisPLB2009}
J. Ellis, N.E. Mavromatos and D.V. Nanopoulos,
Phys. Lett. B674 (2009) 83
 [arXiv:0901.4052]


\bibitem{hessREVIEW} J. Bolmont and A. Jacholkowska,
arXiv:1007.4954, Adv. Space Res. 47 (2011) 380

\bibitem{nemiro} R.J. Nemiroff, J. Holmes and R. Connolly,
Phys. Rev. Lett. 108 (2012) 231103

\bibitem{magic}
J. Albert {\it et al},
Phys. Lett. B668 (2008) 253

\bibitem{hessPRL2008} F. Aharonian {\it et al},
Phys. Rev. Lett. 101 (2008) 170402

\bibitem{gacDAFNEpiran} G. Amelino-Camelia, D. Guetta and T. Piran,
 arXiv:1303.1826

\bibitem{icecubenature} R. Abbasi {\it et al}.,
Nature
484, 351 (2012).

\bibitem{icecubetesi} N. Whitehorne, {\it A search for high-energy neutrino emission from Gamma-ray Bursts},
Ph.D. thesis (https://docushare.icecube.wisc.edu/dsweb/Get/Document-60879/thesis.pdf);
see in particular figures 6.1 and 6.2 and their captions.

\bibitem{wandeLONG}
D. Wanderman and T. Piran,
MNRAS 406, 1944 (2010)

\bibitem{cowaSHORT}
D. Coward {\it et al}.,
MNRAS 425, 2668 (2012)

\bibitem{dsrds} G. Amelino-Camelia, A. Marciano, M. Matassa and G. Rosati,
arXiv:1206.5315, Phys. Rev. D86 (2012) 124035

\bibitem{cina130427a}
Y.-Z. Fan, P.H.T. Tam, F.-W. Zhang, Y.-F. Liang, H.-N. He, B. Zhou, R.-Z. Yang, Z.-P. Jin and D.-M. Wei, arXiv:1305.1261

\end{thebibliography}
\end{document}